\documentclass[11pt,fleqn]{article}

\usepackage{amsfonts}
\usepackage{amsmath}
\usepackage{amssymb}
\usepackage{mathtools}
\usepackage{mathrsfs}
\usepackage{enumerate}
\usepackage{bbm}
\usepackage{graphicx,epsfig,amsthm,color}
\usepackage{pstricks}
\usepackage{subfig}
\usepackage{graphicx}
\usepackage{units}

\usepackage{caption}
\usepackage{booktabs}
\usepackage{textcomp}
\usepackage{array}
\usepackage{cite}
\usepackage{cancel}

\date{\today}

\newcolumntype{z}[1]{>{\RaggedRight\hspace{0pt}}p{#1}}
\newcolumntype{w}[1]{>{\RaggedRight\hspace{0pt}}p{#1}}
\newcolumntype{v}[1]{>{\Centering\hspace{0pt}}p{#1}}

\def\be{\begin{equation}}
\def\ee{\end{equation}}
\def\bea{\begin{eqnarray}}
\def\eea{\end{eqnarray}}

\def\be{\begin{equation}}
\def\ee{\end{equation}}
\def\bea{\begin{eqnarray}}
\def\eea{\end{eqnarray}}

\def\erp2{{\rm e}^{2\rho}}
\def\erm2{{\rm e}^{-2\rho}}
\def\er4{{\rm e}^{4\rho}}

\def\be{\begin{equation}}
\def\ee{\end{equation}}
\def\bea{\begin{eqnarray}}
\def\eea{\end{eqnarray}}

\def\m0{m_{\nu_{0,i}}}

\def\T0{T_{\nu_0}}

\newcommand{\half}{\frac{1}{2}}

\newcommand{\beqa}{\begin{eqnarray}}
\newcommand{\eeqa}{\end{eqnarray}}
\newcommand{\bpr}{\begin{problem}}
\newcommand{\epr}{\end{problem}}
\newcommand{\bcent}{\begin{center}}
\newcommand{\ecent}{\end{center}}
\newcommand{\bfig}{\begin{figure}}
\newcommand{\efig}{\end{figure}}
\newcommand{\bpc}{\begin{picture}}
\newcommand{\epc}{\end{picture}}

\renewcommand{\and}{A_{0}^{\nu ,D}(s)}

\newcommand{\bee}{\begin{equation}}

\def\beq{\begin{eqnarray}}
\def\eeq{\end{eqnarray}}

\newcommand{\bright}{\begin{flushright}}
\newcommand{\eright}{\end{flushright}}
\newcommand{\bminip}{\begin{minipage}}
\newcommand{\eminip}{\end{minipage}}

\DeclareCaptionLabelSeparator{mysep}{\hspace{3pt}:\hspace{3pt}}
\DeclareCaptionLabelFormat{mypiccap}{Fig.\hspace{3pt}{#2}}
\DeclareCaptionLabelFormat{mytabcap}{Tab.\hspace{3pt}{#2}}

\begin{document}

\date{}
\title{
\vskip 2cm {\bf\huge The cosmological constant problem in heterotic-M-theory and the orbifold of time}\\[0.8cm]}

\author{{\sc\normalsize Andrea Zanzi\footnote{E-mail: andrea.zanzi@unife.it} \!
\!}\\[1cm]
{\normalsize Dipartimento di Fisica e Scienze della Terra, Universit\'a di Ferrara - Italy}\\
[1cm]
}
\maketitle \thispagestyle{empty}
\begin{abstract}
{Chameleon fields are quantum fields with an increasing mass as a function of the matter density of the environment. Recently chameleon fields have been exploited to solve the cosmological constant problem in the Modified Fujii's Model - MFM [Phys Rev D82 (2010) 044006]. However, gravity has been treated basically at a semiclassical level in that paper. In this article the stringy origin of the MFM is further discussed: as we will see, the MFM can be obtained from heterotic-M-theory. Consequently, a quantum description of gravity is obtained and the theory is finite because we choose the string mass as our UV cut-off. This stringy origin of the MFM creates stronger theoretical grounds for our solution to the cosmological constant problem.  In our  analysis, time will be compactified on a $S^1/Z_2$ orbifold and this peculiar compactification of time has a number of consequences. For example, as we will see, quantum gravity and a quantum gauge theory are actually the same theory in the sense that gravity is the time-evolution of a gauge theory. This might be the key to obtain a non-approximated stabilizing potential for the dilaton in the string frame. In this paper we will further discuss the non-equivalence of different conformal frames at the quantum level. As we will see, in our approach we use basically a unique conformal frame: the frame where the masses of particles are field dependent. A word of caution is necessary: we do not take into account  massive string states and IR divergences. }
\end{abstract}

\clearpage

\newpage

\setcounter{equation}{0}
\section{Introduction}

One of the crucial problems in modern cosmology is the cosmological constant (CC) one \cite{Weinberg:1988cp}. Recently, a solution to the CC problem has been discussed with the help of chameleon fields in \cite{Zanzi:2010rs}. Chameleon fields \cite{Khoury:2003rn, Khoury:2003aq} are quantum fields, typically scalar, with a mass which is an increasing function of the matter density. Therefore, locally, these fields are heavy while, globally (on cosmological distances) or whenever the matter density is small, the chameleon is light. This peculiar density-dependent mass justifies the name ''chameleon''. For reviews about chameleon fields the reader is referred to \cite{Mota:2006fz,
Waterhouse:2006wv, Weltman:2008ll, Khoury:2013yy}.

In reference \cite{Zanzi:2010rs}, gravity is treated basically at a semiclassical level in the framework of the Modified Fujii's Model (MFM), but it is common knowledge that the CC problem is really acute only in the quantum gravity regime. Is it possible to provide a quantum description of gravitation in the MFM? One step forward has been done in \cite{Zanzi:2014twa} at the level of the effective action with the formulation of a Chameleonic Equivalence Principle (CEP): with the help of chameleon fields it is possible to show that quantum gravitation is equivalent to a conformal anomaly in the MFM \cite{Zanzi:2014twa}. 

One of the purposes of this article is to discuss the quantum description of gravity in the UV completion of the MFM. In particular, we will show that the MFM is obtained from heterotic-M-theory. This connection with the string provides stronger theoretical grounds for our solution to the CC problem in the MFM. Another good news is that the theory is UV finite because we identify the string mass as our UV cut-off. One word of caution is necessary: we will not take into account massive string states and IR divergences.

Formally we have two conformal frames in the MFM, but, as we will see, there is actually a unique conformal frame: the frame where particles' masses are field-dependent.  We will call {\it string frame} (S-frame) the conformal frame where a non-minimal coupling term (dilaton-dilaton-curvature) is present. One essential element of the MFM is a conformal transformation from the S-frame to the Einstein frame (E-frame). Interestingly, a shift of the E-frame dilaton $\sigma$ is related to a shift in the amount of scale invariance in the Einstein frame. This comment is the crucial property exploited in \cite{Zanzi:2010rs} to solve the cosmological constant (CC) problem. The role of conformal transformations in the MFM has been discussed in connection to other problems, for example (A) quantum gravity and the collapse of the wave function in quantum mechanics \cite{Zanzi:2014twa}, (B) solar physics \cite{Zanzi:2014aia}. 
During this analysis we will analyze once again the non-equivalence of different conformal frames at the quantum level proposed in \cite{Zanzi:2010rs}.

Our analysis of the MFM will be developed exploiting a peculiar compactification of time: time will be compactified on a $S^1/Z_2$ orbifold. There are a number of consequences of this compactification. For example, as we will see, gravity is the time evolution of a gauge theory. In this paper we will sometimes refer to this gauge theory as ''strong interaction'' but it is not yet clear whether this theory is really QCD or not. Let us further illustrate these issues.\\
How can we identify gravity with a quantum gauge interaction?
In order to identify strong interaction with gravity, we exploit holography and AdS/CFT \cite{Maldacena:1997re} (for a recent book about AdS/CFT see \cite{Natsuume:2014sfa}). Indeed, in AdS/CFT, gravity in $N$ dimensions is dual to a gauge theory in $N-1$ dimensions. This shift $(N-1) \rightarrow N$ in the number of dimensions is the signature of the holographic nature of AdS/CFT. The Maldacena's conjecture is useful in the MFM, because, as we will discuss in this article, we will exchange space with time and, in this way, the holographic shift $(N-1) \rightarrow N$ is simply due to the time dimension. In other words, AdS/CFT guarantees that gravity is dual to a gauge interaction, but, in our paper, the holographic shift $(N-1) \rightarrow N$ is due to time and, hence, we really identify gravity with strong interaction, in the sense that, as already mentioned above, {\it quantum gravity is the time-evolution of a gauge interaction}.

As far as the organization of this paper is concerned, in section \ref{model} we summarize some useful results already discussed in the literature. The remaining sections contain the original contributions of this article.   In section \ref{four} we analyze the non-equivalence of different conformal frames at the quantum level. Section \ref{string} provides the link between the MFM and heterotic-M-theory.  The last section contains some concluding remarks.

\setcounter{equation}{0}
\section{Some useful results from the literature}
\label{model}

\subsection{The MFM}

We write the string frame lagrangian as (the gauge part is not written explicitly but it is present in the theory)
\beq {\cal L}={\cal L}_{SI} + {\cal L}_{SB}, \label{Ltotale}\eeq where the
scale-invariant (SI) part of the Lagrangian is given by (see also \cite{Zanzi:2010rs}):

\begin{equation}
{\cal L}_{\rm SI}=\sqrt{-g}\left( \half \xi\phi^2 R -
    \half\epsilon g^{\mu\nu}\partial_{\mu}\phi\partial_{\nu}\phi -\half g^{\mu\nu}\partial_\mu\Phi \partial_\nu\Phi
    - \frac{1}{4} f \phi^2\Phi^2 - \frac{\lambda_{\Phi}}{4!} \Phi^4 - \frac{\lambda_{\phi}}{4!} \phi^4
    \right).
\label{bsl1-96}
\end{equation}
$\Phi$ is a scalar field representative of matter fields,
$\epsilon=-1$, $\left( 6+\epsilon\xi^{-1} \right)\equiv
\zeta^{-2}\simeq 1$, $f<0$ and $\lambda_{\Phi}>0$.
One may write also terms like $\phi^3 \Phi$ and $\phi \Phi^3$ which are multiplied by dimensionless couplings. However
we will not include these terms in the lagrangian. Happily, a $\phi^4$ term does not clash with the solution to the CC problem of reference \cite{Zanzi:2010rs}, because the renormalized Planck mass in the IR region is an exponentially decreasing function of $\sigma$ (see also \cite{Zanzi:2012du}).

To proceed further, let us discuss the symmetry breaking Lagrangian
${\cal L_{SB}}$, which is supposed to contain non-scale-invariant terms,
in particular, a stabilizing potential for $\phi$ in the
S-frame with a positive vacuum energy and no fine-tuned parameters. For this reason we write:
\bea
V_{SB}=\frac{A}{\phi^2}+B+\frac{C}{\phi},
\label{newcasimir}
\eea 
where $A$, $B$ and $C$ are constant parameters. As we will see later in this article, \ref{newcasimir} is obtained through Casimir energy (see also \cite{Zanzi:2012bf} for a related discussion).

Now we move to the E-frame. It has been shown \cite{Zanzi:2010rs} that in the E-frame the dilaton is a chameleon in this model and the E-frame (renormalized) vacuum energy is small in the IR region including all quantum corrections and without fine-tuning. Here is the E-frame lagrangian \cite{Zanzi:2010rs}:
 
 \begin{equation}
{\cal L}_{*}=\sqrt{-g_*}\left( \frac{1}{2} R_* -
    \frac{1}{2} g^{\mu\nu}_*\partial_{\mu}\sigma\partial_{\nu}\sigma + {\cal L}_{* matter}
    \right),
\label{eframe}
\end{equation}
where ${\cal L}_{* matter}$ turns out to be
\begin{equation}
{\cal L}_{* matter}= -
    \frac{1}{2} g^{\mu\nu}_* D_{\mu}\Phi_* D_{\nu} \Phi_* - e^{2 \frac{d-2}{d-1} \zeta
    \sigma} (\xi^{-1} \frac{f}{4} M_p^2 \Phi_*^2+ \frac{\lambda_{\Phi}}{4!} \Phi_*^4)
\label{lmatter}
\end{equation}
and $D_{\mu}=\partial_{\mu}+ \zeta \partial_{\mu} \sigma$. $\sigma$ is the E-frame dilaton. The scale non-invariant lagrangian of the S-frame corresponds to a constant contribution to the vacuum energy (the fields are stabilized) and, hence, after the conformal transformation to the E-frame, an exponential potential is obtained. The $\phi^4$-term is mapped into  $M_p^4$ and the Planck mass is switched off in the IR. Quantum calculations in the E-frame show the presence of a conformal anomaly \cite{Zanzi:2010rs, Fujii:2002sb} that induces a direct coupling dilaton-matter.

\subsection{Chameleonic Equivalence Principle}

In \cite{Zanzi:2014twa} a Chameleonic Equivalence Principle (CEP) has been formulated in the framework of the MFM. This principle tells us basically two things:
\begin{itemize}
\item All the ground states of the MFM can be connected to each other through a proper conformal transformation.
\item If, exploiting a conformal transformation, we map one ground state of the theory to another one characterized by a different amount of conformal symmetry, then an additional term in the form of a conformal anomaly must be included in the field equations and this conformal anomaly is equivalent to the quantum gravitational field.
\end{itemize}
 
 Remarkably, this principle is a guideline towards the quantum description of gravity in the effective action. A dictionary between General Relativity (GR) and chameleonic quantum gravity can be formulated and it is summarized in Table 1.
 
\begin{table}[t]
\center
\begin{tabular}{| c|c |c | }
\hline
             & GR  & MFM  \\
\hline

Initial observer    & Inertial frame & Conformal ground state     \\
Final observer  & Non-inertial frame  & Non conformal ground state     \\
Transformation & General coordinate transformation & Conformal transformation \\
Gravitational field description & Metric and connection & Conformal anomaly  \\
Equivalence Principle        & Einstein's Equivalence Principle  & CEP     \\
\hline
\end{tabular}\\[1cm]
\caption{Dictionary between GR and chameleonic quantum gravity. A mapping from an initial observer to a final one is considered. During the mapping some ''extra-terms'' must be added in the field equations and, in harmony with the Equivalence Principle of the theory, they describe the gravitational field.}. 
\label{tab1}
\end{table}


\subsection{Heterotic-M-theory in 5 dimensions}

This paragraph will briefly touch upon the 5D limit of heterotic-M-theory. The reader is referred to \cite{Lukas:1998tt, Brax:2002nt} for more details.

Heterotic theory in weak coupling is a 10 dimensional theory, but in strong coupling an additional dimension is present. This 11-dimensional scenario admits, at intermediate energies, a 5D description. The 5th dimension is compactified over an $S^1/Z_2$ orbifold. The bosonic sector of the 5D action is given by:
\bea
S=\frac{1}{2k_5^2} \int d^5 x \sqrt{-g_5} [R + (\partial C)^2 +U],
\eea
where $C$ is the bulk dilaton and $U$ is a bulk potential.
There are two branes located at the fixed points $z_1$ and $z_2$ of the orbifold and described by the actions
\bea
S_{brane1}=-\frac{3}{2k_5^2} \int d^5 x \sqrt{-g_5} U_B \delta(z_1)
\eea
and
\bea
S_{brane2}=+\frac{3}{2k_5^2} \int d^5 x \sqrt{-g_5} U_B \delta(z_2).
\eea
Moreover there is a Gibbons-Hawking term
\bea
S_{GH}= \frac{1}{k_5^2} \int d^5 x \sqrt{-g_4} K.
\eea
If no fields other than the bulk scalar field are present and no SUSY breaking is considered, then we are in a BPS configuration where the bulk equations agree with the boundary conditions. The result is that there is a no-force condition for the branes and $U_B$ is just a superpotential.
If we make the ansatz 
\bea
ds^2=dz^2+a(z)^2 g_{\mu\nu}dx^\mu dx^\nu
\eea

for the line element in 5D and if we use an exponential superpotential of the form ($k$ is constant)
\bea
U_B=4ke^{\alpha C}, \alpha \in \mathbb{R}
\eea
as suggested by supergravity, then the BPS equations (the bulk equations) can be solved. The solution for the dilaton is
\bea
C(z)=-\frac{1}{\alpha}ln(1-4k \alpha^2 z)
\eea
while for the scale factor is
\bea
a(z)=(1-4k \alpha^2 z)^{1/4\alpha^2}.
\eea
In the limit of small $\alpha$ we obtain an exponential scale factor 
\bea
a(z)=e^{-kz}.
\label{exp}
\eea
There is a naked singularity in the bulk (which is expected to be resolved when the full string theory description is taken into account). The singularity presents itself at $z=1/(4k \alpha^2)$ and, as we will see, it is related to the vanishing of the volume of the remaining 6 extradimensions. The singularity is screened by one of the two branes (the ''hidden brane''). The remaining brane can host the standard model particles and hence will be called the ''visible brane''.

A more realistic model can be obtained by putting matter on the branes and detuning the branes' tension. In this case two scalar degrees of freedom will parametrize the positions of the two branes. In particular, from this bosonic sector of the theory we have two moduli in the effective action in the E-frame: a dilaton $Q$ (parametrizing the position  of the center of mass of the two branes) and a radion\footnote{Do not confuse the R of radion with the one representing curvature.} $R$ (parametrizing the distance between the two branes).
$Q$ and $R$ are related to the branes' position by a couple of redefinitions. The first redefinition is
\begin{eqnarray}
\tilde \phi^2 &=& \left(1 - 4k\alpha^2 z_{visible} \right)^{2\beta}, \label{posia1}\\
\tilde \lambda^2 &=& \left(1-4k\alpha^2 z_{hidden}\right)^{2\beta} \label{posia2},
\end{eqnarray}
with
\begin{equation}
\beta = \frac{2\alpha^2 + 1}{4\alpha^2}.
\end{equation}
The second redefinition is
\begin{eqnarray}
\tilde \phi &=& Q \cosh R, \label{posib1} \\
\tilde \lambda &=& Q \sinh R \label{posib2}.
\end{eqnarray}

This 5D braneworld model can be mapped into a 5D bag exploiting a conformal transformation \cite{Zanzi:2006xr}. In particular, the naked singularity is mapped into the centre of the bag. The branes are mapped into $S^4$ spaces which are the boundaries of the bag. In other words, we obtain a spherical potential well with two $S^4$ boundaries. In the effective action we have a bi-scalar-tensor theory of gravitation (we have two moduli: $Q$ and $R$). We follow the analysis of \cite{Brax:2004ym} and, therefore, we assume that the radion is a chameleon field with a large (small) value in the IR (UV) region. This means that near the big-bang we have the hidden brane close to the singularity and we are left with a single modulus (the dilaton). In the MFM we always work with one single modulus: the dilaton $\sigma$. However, in heterotic-M-theory in 5D we have two moduli, namely $Q$ and $R$. This mismatch in the number of moduli we discuss is not a problem, because both moduli, the dilaton and also the radion, are chameleon fields. Consequently, once the matter density is fixed, the values of both fields are determined. In this way we are led to a scenario where the cosmological evolution is parameterized by a single modulus related to the matter density: this is what we call ''the dilaton $\sigma$''. For this reason, in this paper we will not discuss the radion anymore. Once the hidden brane is near the singularity, the stabilization of the bag radius corresponds to the stabilization of the dilaton and it can be obtained exploiting Casimir energy of bulk fields (see \cite{Zanzi:2006xr}).

\setcounter{equation}{0}
\section{Conformal frames in the MFM}
\label{four}

In reference \cite{Zanzi:2010rs} a non-equivalence of different conformal frames at the quantum level has been pointed out in the framework of the MFM. We further elaborate on this concept exploiting a useful analogy with the relativity of time in Special Relativity (SR). The interested reader can find in references \cite{Steinwachs:2013ama, Kamenshchik:2014waa} further results about non-equivalent conformal frames. 

In the MFM an observer is linked to his local environment and, moreover, we know that in chameleonic quantum gravity the conformal transformations replace general coordinate transformations of GR (see \cite{Zanzi:2014twa, Zanzi:2015kha} and table 1). For these reasons we link the observer to his chameleonic ground state. 
If the S-frame observer related to the stabilized $\phi$ dilaton compares the two cosmological constants in the two frames, he will find that 
\bea
\Lambda_E=K \Lambda_S
\eea
where $\Lambda_S$ is the S-frame CC and $\Lambda_E$ is the E-frame CC. In this formula $K<<1$.
If the E-frame observer related to a stabilized $\sigma$ compares the two CC in the two frames, he will find that
\bea
\Lambda_S=(1/K) \Lambda_E.
\eea
Since the two ''relativity factors'' are not the same in the two formulas we mentioned last, we say, in this sense, that the two conformal frames are not equivalent at the quantum level.

Interestingly, the non-equivalence of different conformal frames at the quantum level of the MFM is related to the gauge-fixing for gravity. Indeed, in the MFM the scale invariant sector of the lagrangian plays the role of the gauge invariant lagrangian in non-abelian gauge theories, while the symmetry breaking lagrangian of the MFM (namely the stabilizing S-frame potential) plays the role of the extra-term to the effective lagrangian of non-abelian gauge theories, namely the non-gauge-invariant term of the Faddeev-Popov-De Witt method. Once the S-frame dilaton is stabilized, the vacuum energy is large and the conformal invariance of the theory is broken. One might wonder whether this non-equivalence is only an artifact of the gauge-fixing procedure. However, physical quantities cannot depend on the gauge choice. In the MFM the gauge fixing (namely the S-frame dilaton stabilization) provides a constant mass scale in the theory (the unrenormalized Planck mass) and the result of a measurement of the E-frame CC is a number, namely the ratio between the CC-mass scale and the unit mass (the unrenormalized Planck mass). This ratio is really running with $\sigma$, the effect is observable and it is not a gauge artifact. In other words, $K<<1$ is not a gauge artifact. The fact that we fixed the gauge is important for physics, but we are free to choose the gauge in the way we prefer. We will further analyze the theoretical origin of the unrenormalized Planck mass in section \ref{string}.

Is it possible to say that in the MFM the S-frame and the E-frame are non-equivalent at the quantum level? In reference \cite{Zanzi:2010rs} the answer was affirmative, but we point out that this conclusion is related to the (very reasonable) assumption that a stabilized scalar field and its vacuum energy do not evolve with time. Indeed, in this case, the CC in the S-frame is fixed (to a large value) after the stabilization of the dilaton has taken place. If we perform a conformal transformation to the E-frame, the CC will be small. The final result is that the two frames are non-equivalent because the two relativity factors are different.

However, there is also another possibility. Recently, climbing scalars have been discussed in string theory (see for example \cite{Dudas:2010gi} and related references). Let us suppose for a moment that after the stabilization in the S-frame has taken place, the S-frame dilaton climbs up the potential in the MFM. In this case we would have a running S-frame dilaton which gets fixed and then restarts its evolution once again. As already mentioned above, the fact that we fix the gauge is important for quantum gravity and cannot be neglected. In other words, it is our intention to construct a model which takes into account the fact that the gauge has been fixed. In order to obtain this result, we can change the dynamical variable representing the dilaton (we use $\sigma$ instead of $\phi$). In this way $<\phi>=constant$ is a built-in information in the model and, naturally, we cannot use $\phi$ as our dynamical variable anymore: now the dynamical evolution (climbing included) of the {\it same} field is described by $\sigma$. The conformal transformation of the metric is introduced in order to have a canonically normalized kinetic term for the new variable $\sigma$. In reference \cite{Fujii:2002sb, Fujii:2003pa}, the lagrangian of the model has been considered by Fujii, but he did not fix the gauge and he obtained a model with field-dependent masses in the S-frame and constant masses in the E-frame: he had two different frames. On the contrary, if we fix the gauge, we obtain field-dependent masses not only in the S-frame but also in the E-frame. Indeed, in the S-frame the ''mass term'' for particles is $\phi^2 \Phi^2$ (and $\phi$ has evolved until the gauge fixing) while in the E-frame the masses of particles are planckian, the Planck mass is renormalized and, hence, it is a non-trivial function of $\sigma$. The physical masses are now dynamical with respect to a reference mass which is constant and this constant is related to the gauge fixing. In this way we are in a unique frame before and after the conformal transformation: the frame where the masses are field-dependent. The unit mass (our ''gauge'') is the unrenormalized Planck mass (namely the Planck mass at $\sigma=0$), this is a constant mass and we will analyze its connection to $k_{11}$ of heterotic-M-theory in section \ref{string}. Summarizing, if we accept the possibility that a stabilized S-frame dilaton can restart its evolution climbing its potential in the MFM, then we can formulate our results in a unique conformal frame and the two different cosmological constants are simply the values of the vacuum energy in two different instants of time.

\setcounter{equation}{0}
\section{The MFM from string theory}
\label{string}

The purpose of this section is to show that the MFM lagrangian can be obtained from string theory. The theoretical framework will be heterotic-M-theory. At intermediate energies this 11D Horava-Witten \cite{Horava:1996ma} theory is described by the 5D model discussed above where the fifth dimension is compactified on a orbifold $S^1/Z_2$. Let us see how the MFM lagrangian can emerge from this 5D set-up.

\setcounter{equation}{0}
\subsection{The scale invariant lagrangian from string theory}
\label{SIL}

The aim of this section is to justify the scale invariant sector of the MFM from string theory. Since the term $\phi^2 \Phi^2$ is the scale invariant counterpart of a mass term, we will start our analysis from the Flavor problem. As already mentioned above, the theoretical framework we are going to exploit is heterotic-M-theory and, in particular, its intermediate energy scenario given by a 5D bulk with the fifth dimension orbifolded on $S^1/Z_2$ and two boundary branes (the visible brane and the hidden one) at the fixed points of the orbifold. There is also a naked singularity in the bulk, but it is screened by one of the branes (the brane cuts a slice of the 5D space and the naked singularity is beyond the brane). For a detailed discussion of this model the reader is referred to \cite{Zanzi:2006xr} and references therein. Now let us come back to the Flavor problem.

In the Standard Model (SM), all the fermion masses and mixing angles are free (unpredicted) parameters. Moreover, a huge variety of fermion masses is present: from the MeV region of the electron mass to the (almost) two hundred GeV of the top quark mass (and we are forgetting about neutrinos!). In other words, a ''Flavor Theory'' is completely missing in the SM (see Course 1 of \cite{Kazakov:2006kp} for an introduction to flavor physics and grand unification).

In \cite{Zanzi:2006xr}, a mechanism to generate small Dirac neutrino masses has been discussed in the framework of heterotic-M-theory. In this paragraph we extend that analysis to {\it all} the fermions of the SM. We start considering $N$ identical copies of the SM (see also \cite{Dvali:2007hz}) and $N+1$ branes in the 5D bulk. This multiple 3-branes scenario can be obtained from heterotic-M-theory considering M5 branes wrapped on 2-cycles. We have the hidden brane (we call it the $N+1$-th brane and we imagine it to be close to the naked singularity) and also a stack of $N$ branes, well-separated from the hidden one, where we localize the various SMs. Let us consider for example the j-th brane ($1\leq j\leq N$): here we localize all the particles of the $j$-th copy of the SM with one exception, namely, we put in the bulk a massive Dirac fermion whose right-handed zero mode will be localized close to the singularity (at large redshift). This right handed zero mode will have a non-trivial overlap with the $j$-th brane and it will provide the right-handed component of a certain fermion of the SM (for example the electron). We repeat this construction for the remaining fermions of the SM. In other words, if we consider the $j+1$-th brane, all the particles of the $j+1$-th copy of the SM will be localized on that brane with the exception of one massive bulk Dirac fermion whose right handed zero mode will provide the right-handed component of another particle ({\it not} the electron, but for example the muon) of {\it another copy} of the SM. In this way, we are led to a scenario similar to the Universal Extra Dimension proposal (see \cite{Servant:2002aq}). 

The careful reader may wonder whether the presence of SM fields in the bulk is acceptable in this model. We elaborate this point discussing two issues:\\
1) {\it Gauge fields in the bulk}. SM fields in the bulk are sources of gauge fields and, therefore, the SM gauge group must be in the bulk in our proposal. The presence of gauge fields is rather natural in our scenario, because we have an extended (N=2) gauged supergravity theory in the bulk and, consequently, a large number of bulk vector fields is present. \\
2) {\it The size of the bulk}. The SM fields can be included in the bulk granted that the size of the bulk is not too large. If the extra dimensions are of inverse-TeV size, we can have the SM in the bulk. However, a much larger bulk cannot tolerate the presence of SM fields. For a discussion of one possible choice of parameters of the theory, the reader is referred to section \ref{param}.\\

The fermions in the bulk are massive, the 5D mass term is of the form $M_5 \bar{\Psi}\Psi$ and, hence, it does not clash with the SM gauge group. These masses are similar to each other and not fine-tuned. Every bulk fermion stabilizes through Casimir energy (see \cite{Zanzi:2006xr}) the corresponding brane, granted that we impose proper boundary conditions such that no fermion current is lost through the brane (MIT boundary conditions). The Casimir contribution guarantees the formation of extradimensional condensates of SM fields, namely, gauge singlet scalar fields which are representative of pairs of SM fermions. Following the neutrino case discussed in \cite{Zanzi:2012bf}, we suggest to exploit these condensates as composite Dark Matter in our model. In the absence of fine tuning, the various branes are close to each other in the bulk and well-separated from the hidden brane. In this way, we are led to a multi-scalar-tensor theory with constant masses for matter fields after the Casimir stabilization of branes has taken place. If we call $\phi$ (squared) the object which multiplies the curvature in 4D in the S-frame, we can connect this scenario to the single-field analysis of the MFM.
The massless zero modes of the various bulk fermions are treated as in \cite{Zanzi:2006xr}. Hence, they are varying {\it rapidly} as a function of the 5th dimension and they are localized towards the hidden brane. The dilaton is related to the Higgs field in our model and, therefore, the value of the Higgs on the various branes is obtained from the dilatonic profile in the bulk.
This point we mentioned last requires some additional comments. The dilaton is a gauge singlet while the Higgs boson is not. Hence a direct identification of the two scalar fields is forbidden. In our model we include the Higgs doublet $H$ with the same lagrangian of the $\Phi$ field. Consequently, when we minimize the effective potential of the MFM, we obtain $H \propto \phi$. In this sense we will identify the dilaton with the Higgs field in the MFM.

 Happily, the 5D dilaton is a decreasing function of the distance from the singularity and, if we consider non-fine-tuned values for the fields, a small expectation value of the dilaton can be obtained naturally. From the Moduli Space Approximation analysis discussed in the literature (see for example \cite{Zanzi:2006xr}), we know that the bulk dilaton profile $C(z)$ varies {\it slowly} as a function of the 5-th coordinate, in particular we have:
\bea 
C=-\frac{1}{\alpha}ln(1-4k\alpha^2 z),
\label{dilC}
\eea
where $\alpha$ is a small number and $k$ is a constant (see formula 2.14 of \cite{Zanzi:2006xr}). 

To proceed further, we ask: what is the mass of, for example, the electron? The mass of the electron in the S-frame is the sum of the electronic masses, $m_i$, of the various copies of the SM. All the $m_i$ are small because in the absence of fine tuning the bulk dilaton (which ''plays the role'' of the Higgs field in the sense described above) is almost vanishing, but we have one exception: one (and only one) of these $m_i$ will be non-negligible because it will be the result of the overlap with the right-handed bulk zero mode of the electron and we know that this zero mode varies rapidly as a function of the 5-th coordinate (see \cite{Zanzi:2006xr}). This is the main advantage of this approach to the flavor problem: a small change in the position of a brane in the bulk corresponds to a large variation of the related fermion mass in 4D. In other words, the fermion masses in 4D are telling us the positions of the branes in the bulk. The reader may wonder where are we supposed to live, namely, in which brane. The answer is that we live in the E-frame\footnote{As already mentioned above, we can summarize our analysis in a unique frame granted that the dilaton climbs up the potential after the gauge fixing.}, but we will say more about this many-branes set-up when we will compactify the time dimension on an orbifold. 

As already mentioned above, if we call $\phi$ (squared) the field which multiplies the 4D curvature in the S-frame, we can connect this flavor model to the single field analysis of the MFM. Remarkably, the brane construction described in this paragraph will produce mass terms in the S-frame of the form ''dilaton-fermion-fermion'', which is the fermionic counterpart of the $f \phi^2 \Phi^2$ term in the S-frame of \ref{bsl1-96}. We warn the reader that, at this stage, the sign of $f$ is inserted by-hand. Moreover, the $\Phi^4$-term can be obtained through a 1-loop calculation starting from the vertex $f\phi^2\Phi^2$. Similar 1-loop diagrams can generate also a $\phi^4$ term in the lagrangian and this term does not clash with the solution to the CC problem, because the renormalized Planck mass in the IR region is an exponentially decreasing function of $\sigma$ (see also \cite{Zanzi:2012du}). These comments are supporting a stronger connection between the lagrangian we exploited to solve the CC problem and string theory, however, the stabilizing dilatonic potential in the S-frame has been connected to Casimir energy in \cite{Zanzi:2012bf} granted that some assumptions are accepted. We further elaborate on the stringy nature of the S-frame dilatonic potential in the next paragraph.

To a certain extent, this approach to the flavor problem is similar to \cite{ArkaniHamed:1999dc, Shadmi:2011uy}. A phenomenological analysis of this approach to the flavor problem is required.

\subsection{The orbifold of time and the stabilizing potential for the dilaton}
\label{gf}

We start pointing out that the time evolution of $\sigma$ corresponds to a RG running from UV to IR. In particular, the UV region is the small $\sigma$ region and the stabilized dilaton in the S-frame corresponds to a configuration of far past on cosmological time scales. In other words, the chameleon mechanism makes indistinguishable the time evolution of $\sigma$ from the RG running. After all, this is not surprising: in AdS/CFT a 5D translation is mapped into a 4D dilatation and in heterotic-M-theory the bulk dilaton is a monotonic function of the fifth coordinate $z$.

Now we will explore the potential identification of gravity with a 4D quantum gauge theory on the boundary of the 5D bulk. As already mentioned above, it remains to be seen whether this 4D theory is really QCD or not. To proceed further, we will exchange space with time. This is a well-known trick in string theory and it can be exploited with Euclidean metrics (see for example \cite{Tong:2009np}). In the MFM, the Casimir calculation exploited in our flavor theory is based on a Wick rotation and the time direction is compactified on a circle. To be more precise, in the Casimir calculation (see also \cite{Zanzi:2006xr}),  the 3-branes and the time are mapped by the conformal transformation into $S^4$ spaces (the fourth dimension is the Wick-rotated time). The naked singularity is mapped into the center of the ball. However a conformal transformation cannot change topology and, consequently, a singularity which is screened by one brane in the S-frame will remain screened by one $S^4$ space into the bag frame. We infer that one of the $S^4$ spaces is very close to the singularity, it is the horizon of the black hole and in this sense the naked singularity is removed from the physical region. Hence, the 5D ball (i.e. the bag) is actually a 5D spherical potential well with various boundaries (the $S^4$ spaces) while the fifth dimension of the Horava-Witten set-up is mapped into the radial dimension of the well.

Let us exchange space with time and, in particular, our intention is to consider the fifth dimension as the time coordinate. Let us check whether the metric is Euclidean. The line element is \cite{Zanzi:2006xr}
\bea
ds^2=dz^2+a(z)^2 g_{\mu\nu}dx^{\mu}dx^{\nu}
\eea
where $a(z)=(1-4k\alpha^2 z)^{1/(4 \alpha^2)}$.
As already mentioned in \cite{Zanzi:2006xr}, the metric is conformal to  $I \times  S^4$:
\bea
ds^2=a(z)^2 [dy^2 + g_{\mu\nu}dx^{\mu}dx^{\nu}],
\eea
where $dy=dz/a(z)$.
In this frame the line element in square brackets is Euclidean both in the UV (because locally we can exploit the Einstein's equivalence principle) and in the IR (because our visible universe is a small portion of a large $S^4$ space). Hence, space and time can be exchanged with each other and this procedure is well-established at least in the deep IR and deep UV regions (it remains to be seen whether this procedure can be safely exploited at intermediate length scales and we will not discuss this issue in this paper). Therefore, let us consider the fifth dimension as our time coordinate. We are led to an orbifold of time. Let us analyze this idea with a useful diagram. We consider an Euclidean plane with cartesian coordinates $XY$. Let us identify the large-$X$ region with the large-$\sigma$ region.  Let us consider the time coordinate compactified on a circle centered in the origin of the cartesian frame. Before we impose the $Z_2$ parity the upper half of the circle is separated from the lower half and we can parametrize the position of a point on the circle with the standard angle exploited in trigonometry (let us call it $\theta$). Hence $\theta=0$ corresponds to the intersection of the circle with the positive $X$-axis. $\theta$ is increasing when we move counterclockwise along the circle. We choose the origin of time at $\theta=\pi$ rad (where the naked singularity is located, namely, as we will see, where scale invariance is spontaneously broken). With this choice, $\theta=0$ can be identified with the beginning of a pre-big-bang (PBB) phase where $\sigma$ is large, the time is negative and the heterotic (critical) string is weakly coupled. Let us follow the time evolution of the system\footnote{A detailed cosmological discussion of the link between the pre-big-bang scenario and the MFM will be analyzed in a future work. The reader interested in pre-big-bang physics is referred to \cite{Gasperini:1992em, Gasperini:2002bn} and references therein.}. At the end of the pre-big bang phase, $\theta$ is almost $\pi$ rad, the theory is strongly coupled and the time is negative. At the beginning of the post big bang phase the time is positive, $\theta$ is slightly larger than $\pi$ rad and the theory is strongly coupled. The universe at low redshift is described by $\theta$ slightly smaller than $2 \pi$ rad, the theory is weakly coupled and the time is large and positive.
To proceed further, we impose the $Z_2$ parity of the orbifold: we identify $t$ and $-t$. Remarkably, the $Z_2$ orbifold parity identifies our low redshift universe with the universe at the beginning of the PBB phase. 

Let us come back to the stabilizing potential for the dilaton and to the identification of gravity with the 4D gauge theory. This identification might give us an {\it exact}\footnote{In QCD the stabilization of the hadronic radius based on the MIT bag model is related to non-perturbative physics. Indeed, the bag radius $R_{BAG}$ is related to $\Lambda_{QCD}$ ($\hbar c \simeq \Lambda_{QCD} R_{BAG}$). Therefore, the stabilization of the bag radius in the MIT bag model takes into account all the quantum corrections in powers of $\alpha_{strong}$ because the theory is non perturbative. If the gauge theory is really QCD, the gauge-fixing potential is exact.} gauge-fixing potential for gravity. In the 5D bulk there is a naked singularity and, therefore, the invariance under translations in 5D is spontaneously broken. It is common knowledge that a translation in 5D corresponds to a RG running in 4D through AdS/CFT and, hence, the singularity corresponds to a spontaneous breaking of scale invariance in 4D. This symmetry breaking is the source of the Planck mass in the MFM \cite{Fujii:2003pa}. In other words, the unrenormalized $M_p$ has been physical when the universe was small and the volume $V$ of the 6 extradimensions was small (indeed the singularity is in the bulk but not on the brane). The volume $V$ is the volume of a 6D manifold fibered (see the next paragraph for more details) along the 11D. The unrenormalized $M_p$ is a mass scale related to the physics inside the singularity (the hidden brane ''fell'' into the singularity) because this mass scale is produced by the spontaneous breaking of scale invariance. This is a curvature singularity and it might be that the physics inside this black hole is described on the boundary of the black hole (i.e. holography). If we follow this holographic idea, the $S^4$ manifold close to the singularity contains the information about the interior of the black-hole. However, this $S^4$ manifold is also a boundary of the ''bulk of the bag'' and we infer that the physics of the bulk is related to the physics of the black hole through holography because the black hole and the 5D bulk have a common boundary.  The unrenormalized Planck mass is constant because it is related to the physics inside the black hole.  

Now to the point. This holographic connection between quantum gauge interaction in 4D and gravity in 5D allows a total identification between gravity and 4D gauge interaction, granted that our fifth dimension is time (i.e. after the exchange of space and time). In other words, after the exchange of space and time, the holographic ''jump'' $(N-1) \rightarrow N$ is due to time and, consequently, the gauge-gravity duality of AdS/CFT becomes, in our proposal, a true identification between gauge interaction and gravity. This point must be clarified.
In AdS/CFT there is a duality between 5D gravity and the 4D gauge theory on the boundary. Let us now exchange the 5th space coordinate with time. Inspired by AdS/CFT let us imagine a renormalizable 4D gauge theory on a 4-brane. Time is orbifolded and all the 4 dimensions of the brane are spacelike (in other words, this 4-brane is basically an S-brane, see \cite{Gutperle:2002ai}). Let us discuss the time-evolution of this 4-brane. This time evolution is indistinguishable from a holographic ''jump'' in the number of dimensions. Hence, inspired by AdS/CFT, we obtain a 5D gravity theory and, therefore, there is no real difference between 5D gravity and the 4D renormalizable gauge theory: {\it gravity is the time-evolution of a 4D gauge theory}. Indeed, let us follow the time evolution of a brane. Our multiple branes scenario can be summarized by the time evolution of a single 4-brane B. Time is quantized, so the number of branes is large but discrete (this quantization of time is analogous to the chronon of \cite{Caldirola:1977fu}). B evolves with time (i.e. it moves along the 5° dimension) and whatever will be the instant of time T that we choose, there will be one corresponding brane B(T), namely the brane considered at the time T. We can extract a 4D theory on the boundary starting from our 5D gravity theory: we must choose one instant of time T. Consequently, 5D gravity at time T is a (renormalizable by assumption) 4D quantum gauge theory.

The issue of renormalizability should be further discussed. The theory is stringy, therefore, it is UV finite and there are no problems of convergence. However, it seems worthwhile pointing out that the identification of gravity with the time evolution of a 4D gauge theory is telling us that even if we remain {\it at the field theory level} there are no problems at the UV because the gravitational theory is renormalizable.

Summarizing, in this model, gravity is the time evolution of a 4D quantum gauge interaction. If we start from 5D gravity and we choose one instant of time we do not simply obtain ''gravity at time T'', but, exploiting AdS/CFT, we obtain the 4D quantum gauge theory on the boundary. Consequently gravity at time T is a 4D quantum gauge theory.

The connection of the lagrangian of the MFM with string theory is based on some assumptions:\\
1) we assume to put the SM fields in the bulk. This assumption is not new and it has already been discussed in the stringy literature.\\
2) We assume a detuning to the dS direction as in \cite{Zanzi:2012bf}. \\
3) We assume that, in heterotic-M-theory, the non minimal coupling term, the dilatonic profile in the 5th dimension and the kinetic terms for the fields are not perturbed by strong coupling corrections. The kinetic terms are assumed to be compatible with the MFM.\\
4) We assume to parametrize the position of the stack of $N$ branes separated from the singularity through one single modulus.\\

We think that the analysis presented in this work regarding the connection of the MFM with string theory is a step forward with respect to reference \cite{Zanzi:2012bf}.

\subsection{Chameleon-induced compactification}
\label{param}

 Let us consider the 11-dimensional action of the model (see \cite{Witten:1996mz}). When we compactify it down to 4 dimensions we can write \cite{Witten:1996mz}:
\bea
G_N=\frac{k_{11}^2}{16 \pi^2 V \rho}
\eea
and also
\bea
\alpha_{GUT}=\frac{(4 \pi^2 k_{11}^2)^{2/3}}{2V}
\eea
where $V$ is the volume of the 6-dimensional space, $\rho$ is the size of the orbifold and $k_{11}$ is the gravitational scale in 11 dimensions.
Neglecting numerical factors we have
\bea
M_p^2 \propto \frac{V\rho}{k_{11}^2}
\eea
and 
\bea
\alpha_{GUT}\propto \frac{k_{11}^{-2/3} \rho}{M_p^2}.
\eea
From this last formula we identify our UV cut-off, namely the string mass, as
\bea
M_S^2 \propto \rho k_{11}^{-2/3},
\eea
in harmony with the results of the literature (see for example \cite{Antoniadis:1999yx}).
In the language of the Moduli Space Approximation \cite{Zanzi:2006xr}, the Planck mass is related to the Q-field and the string mass is related to the radion $R$.
Hence we have two scalar fields parametrizing two different mass scales. However in \cite{Zanzi:2006xr}, when we fix the position of the hidden brane, a single degree of freedom, namely the position of the other brane, is related not only to the separation of the branes (i.e. the radion), but also to the position of the center of mass of the branes (i.e. the dilaton). In this way, this degree of freedom controls both mass scales and this is the situation considered in \cite{Zanzi:2012ha, Zanzi:2012bf}. In this paper we have many branes (one brane close to the singularity near the big bang and a stack of $N$ branes), but we assume to parametrize the position of the stack of branes with a single modulus. As already mentioned above, a more detailed discussion of the cosmological evolution of the branes is left for a future work.

Another comment is necessary. The reader might be worried by the presence of the mass scale related to $k_{11}$ (the only mass scale in M-theory) because it might clash with scale invariance. This problem can be solved if we connect $k_{11}$ to the unrenormalized Planck mass $M_p^{unren}$. Indeed, this mass scale is constant but it does not clash with the restoration of scale invariance in the IR for two reasons: 1) it is related to the physics inside the black-hole and, hence, this physics is screened; 2) it is unrenormalized and, hence, it does not take into account backreaction (see \cite{Zanzi:2010rs}). Therefore we write
\bea
k_{11} \simeq (M_p^{unren})^{-9/2} \simeq (10^{19} GeV)^{-9/2}.
\eea
A scale invariant theory in the IR can be obtained granted that $\rho$ is small enough \footnote{In particular $\rho$ must go to zero faster than $1/V$ otherwise the renormalized Planck mass clashes with the restoration of scale invariance in the IR} and this is natural because in the IR the 11-th dimension is lost. Since $\alpha_{GUT}$ must be switched off in the IR (in harmony with \cite{Zanzi:2010rs}), we understand that the volume $V$ must be large in the IR, indeed we have
\bea
\alpha_{GUT} \propto \frac{k_{11}^{-2/3} \rho}{M_p^2} \propto \rho/(V \rho) =1/V.
\eea
Summarizing, the chameleonic solution to the cosmological constant problem presented in \cite{Zanzi:2010rs} can be embedded in heterotic-M-theory. At the level of the effective lagrangian, the quantum aspects of gravitation are contained in the CEP \cite{Zanzi:2014twa}. The stabilizing potential for the dilaton is exact if we identify gravity with QCD. The string length is chameleonic \cite{Zanzi:2012ha} and the volume $V$ gets large in the IR (this aspect of the model is reminiscent of the DGP scenario \cite{Dvali:2000hr}). The volume $V$ is fibered along the orbifold of time because $V$ is related to the dilaton and the value of the dilaton runs during the cosmological evolution. Let us further elaborate the connection between the string length and the volume $V$.

One of the main problems in string theory is the origin of compactification. The theory is formulated in more than 4 dimensions, and these extradimensions, so far, have not been detected in experiments. A dynamical mechanism for the compactification would be welcome.
As already mentioned in \cite{Zanzi:2012ha}, the string mass (and the string length) is chameleonic in our model. If we extend this property to strings living in higher dimensions, we obtain a compactification induced by chameleon fields.
Let us start considering the gravitational coupling $\alpha_G$. In the weak coupling regime, $\alpha_G$ is an exponential function of $\sigma$ (see also \cite{Zanzi:2012ha}) and, moreover, the Planck mass is an exponential function of $\sigma$. Hence, if we exclude very small values of $\sigma$, an exponential dependence of the string mass on $\sigma$ is expected.
This result can be interpreted through a relation between $\sigma$ and the winding number of strings in extradimensions. This point needs to be clarified. In string theory, it is very common to wrap strings around spatial extradimensions. In our 5D heterotic set up, the 5th dimension is time and we don't wrap strings around time. However, we can imagine that one closed string (heterotic theory contains only closed strings) is wrapped around a spatial extradimension and, moreover, we can imagine (just to make a simple example) that this extradimension is compactified on a circle with a very large radius. Why do we ''see'' only four dimensions in our experiments? We, observers, live in a ground state where conformal symmetry is abundantly broken through the chameleon mechanism. We infer that the string tension is large and, therefore, the string wrapped around the circle is able to reduce the size of the extradimension. Consequently, the winding number is larger than before and the string looks shorter. In other words, the MFM is telling us that the reason why we don't manage to detect extradimensions is that we live in a ground state where conformal symmetry is broken. In this sense, the chameleonic behaviour of conformal symmetry induces the compactification in our model and, as already mentioned above, {\it in the IR the universe is higher dimensional} (a result similar to the analysis of \cite{Dvali:2000hr}). 

How is it possible to have a large bag radius for large values of $\sigma$ if we are in the weak coupling regime and, therefore, the 11th dimension is absent?  In this room the string mass and the Planck mass are large because $\rho$ is large (we are in the non perturbative regime). When we shift $\sigma$ to large values, we recover the weak coupling regime of the theory and, therefore, $\rho$ goes to zero. In other words, on cosmological distances, the size of the orbifold is almost vanishing and the branes at the fixed points are close to each other. On the other hand, the post-big-bang cosmological expansion must increase the bag radius (because the expansion is parametrized by the dilaton which is a monotonic function of the 5th dimension). As already mentioned above, the chameleonic radion runs with time from small values near the big-bang to values of the order of one in our low-redshift universe. Hence, we are led to the following scenario. In this theory, in the IR region, all the branes are very close to each other ($\rho$ is small), but they are also far away from the singularity (because the bag radius is large after the renormalization of the Planck mass is taken into account). With the notations of \cite{Zanzi:2006xr}, in the IR region (and, hence, in the E-frame) we have $(<\tilde \phi>)^2 \simeq (<\tilde \lambda>)^2 >>0$ a configuration compatible with a small Planck mass. If $\rho$ decreases fast enough, we restore scale invariance and we have cosmic strings in the IR like in our papers \cite{Zanzi:2010rs, Zanzi:2012ha}.

The situation outlined above, namely the exponential dependence of the string mass on $\sigma$, has an interesting classical counterpart. Let us suppose that a sailor wants to fix a boat in a harbor. We can imagine that a small cylinder is available at the harbor and the sailor is supposed to wrap a string around it. Typically, sailors don't make knots when they fix their boats. They simply wrap the string several times around the cylinder and then they let the string fall on the floor. The reason is that the string tension increases exponentially with the winding number. Since the exponential function is rapidly increasing, when the winding number is, for example, 5, the boat is already fixed (no knot is necessary).

\subsection{Discussion}

Needless to say, the careful reader might be puzzled by this analysis. For this reason, let us imagine a ''dialog'' between the reader (R) and the author (A):
\\
R. Is it possible to link the MFM to the Randall-Sundrum (RS with two-branes) model \cite{Randall:1999ee}?\\
A. The RS model is linked to the 5D Horava-Witten scenario (like the MFM) but with frozen Calabi-Yau moduli. In the MFM, on the contrary, the volume $V$ is chameleonic. We can retrieve the AdS profile starting from the 5D set-up related to the MFM by performing the limit $\alpha \rightarrow 0$ (see also \cite{Zanzi:2006xr}). \\
R. In the E-frame of the MFM, we have different ground states and different values of the dilaton when we modify the matter density. From the 5-dimensional point of view, this shift of the dilaton is related to a 5D translation. If we exchange space and time, what is the meaning of a translation in the time direction?\\
A. A translation in the time direction means that different 4D ground states in the E-frame are related to different points in the orbifold of time. Indeed, when we switch on the matter density, we switch on the conformal anomaly (i.e. gravity) and, consequently, we obtain an RG running in 4D which corresponds to a translation along the time direction. In other words, when we change the gravitational field, we modify the ''speed of time''. This phenomenon is already present in GR, where it is common knowledge that the ''speed of time'' is related to the gravitational field.\\
R. Globally the metric is not Euclidean because $g_{\mu\nu}$ corresponds to a de Sitter space. How can we exchange space with time?\\
A. It is true that the 4D part of the metric is dS, but we live on a small portion of the $S^4$ space. Therefore, it is true that globally the metric is {\it not} euclidean, but it is euclidean {\it as far as all practical purposes are concerned}.\\
R. The idea that gravity is the time evolution of a gauge theory is generic or not? In other words, would it be possible to discuss this issue in a more generic way?\\
A. The idea that the time evolution of a quantum gauge theory produces gravity can be obtained granted that (1) the compactification manifold is properly chosen and (2) time is exchanged with one spatial direction. Indeed, we can also forget about heterotic theory for a moment and imagine a string theory where gauge interactions are described through Chan-Paton factors of an open string. Let us assume that this open string lives in an $S^4$ space and, in order to make the analysis simpler, let us work in two dimensions (i.e. with an $S^2$ space). One circle is a timelike coordinate and the ''orthogonal'' circle is spacelike: $S^2$ is obtained as a product of two $S^1$ spaces. An open string ''at rest'' describes a two dimensional manifold homeomorphic to a cylinder because it moves along the timelike circle. When we exchange space and time, the timelike circle becomes a spacelike circle and vice-versa. After this space-with-time exchange, the ''cylinder'' is exactly the worldsheet of a closed string: we found gravity. Once again, gravity is the time evolution of a gauge interaction. Remarkably, with this strategy, gravity is a direct result of a non-trivial geometry for the compactification of spacetime: we can also start with a gravity-free theory and the $S^1$ compactification will make the loop with the open string producing gravity. This seems to be basically an induced gravity theory which exploits the geometry of compactification.\\
R. In string theory the trace anomaly produces higher powers of the curvature and these terms are expected to be important in the strong coupling/high energy regime. Where are these contributions in the effective action of the MFM?\\
A. These contributions are included in the Casimir potential. First of all, it is common knowledge that in two dimensions the conformal anomaly is related to the Casimir energy (see for example \cite{Tong:2009np}). Something similar happens in 4D. Indeed, in the E-frame the conformal anomaly stabilizes the dilaton $\sigma$ and brings it back to small values. In $\sigma=0$ the stabilizing contribution of the conformal anomaly is glued with the Casimir potential of the 5D bag. This is in harmony with the chameleonic equivalence principle: we know that the Casimir potential is a quantum gravity calculation (because it can be obtained also for a sterile spinor which interacts only gravitationally) and, therefore, {\it the CEP is telling us that the Casimir potential is a conformal anomaly}. Trace anomalies have been calculated also at all orders in the literature (see for example \cite{Adler:2004qt}) and this might be the key to obtain a non-approximated stabilizing potential for the dilaton in the S-frame. A promising line of development will further analyze this point. We can understand, at least qualitatively, the structure of the Casimir potential as the result of a linear combination of various powers of the curvature. Indeed, if we exclude the possibility of a fast change of the cosmological dilaton with time, it is natural to expect a dependence of the curvature on $\sigma$ of the form $R \simeq e^{-2\zeta \sigma}$. If we make linear combinations of various powers of the curvature and if we (reasonably) assume that the sign of the coefficients of the curvature is not constant, then we combine together increasing and decreasing functions of $\sigma$ and the result is a  stabilizing (Casimir) potential. The situation is reminiscent (at least qualitatively) of Newtonian gravity: when we consider a point particle in the presence of a central potential $1/r$, then we can construct an effective potential where a minimum is obtained by making linear combinations of the standard Newtonian potential with the $1/r^2$ term. \\
R. There are conformal anomalies in quantum field theory and conformal anomalies in string theory. The higher powers of the curvature mentioned above are a stringy effect. Are we sure that the Casimir energy is related to the string?\\
A. In harmony with the compactification induced by a chameleonic string, we can imagine a string wrapped around a spatial dimension. The non-vanishing string length introduces a boundary which corresponds to the boundary of the Casimir effect.  Indeed, as already discussed in the first volume of the Polchinski's string book, we can relate the Casimir energy to the string length. In the UV the string length is short in our model and, consequently, the Casimir energy is large. In other words, the boundary of the Casimir effect is the bag surface (or, to be more precise, the various $S^4$ spaces), but the bag radius is proportional to the string length. In particular, the bag radius will not be very different from the cosmological value of the string length. In this sense, our Casimir is a stringy effect and the anomaly is stringy. One more comment is in order. With this strategy, the Casimir energy is simply an $\alpha'$-correction. Indeed, $\alpha'$-corrections are important when the string length becomes comparable to the curvature scale of spacetime and this is precisely the way in which we support the stringy nature of our Casimir energy. Hence, the Casimir energy is an $\alpha'$-correction.  \\
R. Do we recover weakly coupled heterotic theory in the IR?\\
A. Typically, the string mass is fixed. In our scenario we fix $k_{11}$ while the string mass is moduli-dependent and variable. In this sense, our IR limit is a weakly coupled theory that is {\it not} weakly coupled heterotic string theory.\\
R. An orbifold of time seems to clash with the increasing behaviour of the entropy with time. Moreover, there might be violations of the causality condition and time travels. These considerations seems to rule out an orbifold of time.\\
A. As far as the entropy of the universe is concerned, there is no clash with the orbifold of time, because the entropy of the entire universe (not only the visible universe) is constant in this model. Indeed, whatever will be the thermodynamical transformation that we consider, on cosmological time scales the orbifold of time will make it reversible. If we make a ''film'' of the transformation from the instant $t_1$ to $t_2$, then, after a long enough time, we will obtain the inverse transformation (i.e. the transformation that we obtain playing the film backwards) from $-t_2$ to $-t_1$. The entire universe cannot exchange heat. Therefore, at the cosmological level the generic transformation is adiabatic, reversible and, hence, isentropic. As far as the potential violations of the causality condition or time travels are concerned, we have no answers at the moment. More research efforts are necessary to clarify these issues. At this stage we simply point out that these problems might arise also with a simple $S^1$ compactification of time: they are not necessarily related to the orbifold of time.

\section{Conclusions}

Our paper provides stronger theoretical grounds to the solution to the CC problem of reference \cite{Zanzi:2010rs}. 
Here is a summary of our results.

We showed that the lagrangian of the MFM can be obtained from heterotic-M-theory. Consequently, the theory is UV finite because we identify the string mass with our UV cut-off. During our analysis we exploited a very peculiar compactification of time: time is compactified on a $S^1/Z_2$ orbifold. 

 The quantum theory of gravitation obtained at low energy has some characteristic features that we summarize here:
\begin{itemize}
\item The theory becomes extradimensional in the IR (analogously to the DGP model) and the compactification of the extradimensions is induced by the chameleonic behaviour of the string tension (see also \cite{Zanzi:2012ha}).
\item  We showed that gravity is the time-evolution of a gauge theory and this result depends on the exchange of space with time. This exchange is possible in the deep UV and deep IR regions of our model, it remains to be seen whether it is allowed also at intermediate length scales. 
\end{itemize}

 The non-equivalence of different conformal frames at the quantum level in the MFM has been discussed in connection to the relativity factor of SR. If we accept the possibility that the stabilized S-frame dilaton can restart its evolution climbing up the potential, we can summarize our results working in a unique conformal frame.\\

Needless to say, there are many open issues. We did not take into account massive string states and IR divergences. An interesting line of development will try to better understand the origin of $k_{11}$ in connection to the physics of the extradimensional black-hole.  Another promising line of development is given by climbing scalars in connection to the MFM. As far as phenomenology is concerned, a detailed analysis of these ideas is necessary (for example of our approach to the flavor problem).  As far as the stabilizing potential for the dilaton is concerned, if we identify the gauge theory with QCD then we can claim that the potential obtained through the 5D MIT bag model is exact, but this has not been proved in this paper and remains as an open issue.

\subsection*{Acknowledgements}

I am grateful to Ignatios Antoniadis, Massimo Bianchi, Marco Bochicchio, Maurizio Gasperini, James Gray, Claus Kiefer, Antonio Masiero, Jose Francisco Morales, Massimo Pietroni, Gianfranco Pradisi, Augusto Sagnotti, Gabriele Veneziano for useful conversations. I warmly thank the Galileo Galilei Institute for Theoretical Physics (Florence) for the kind hospitality: part of this work has been developed at the GGI.



\providecommand{\href}[2]{#2}\begingroup\raggedright\endgroup

\end{document}